\documentclass[aps,pra,preprint,superscriptaddress]{revtex4-1}
\usepackage{graphicx}
\usepackage{epsfig}
\usepackage{epstopdf}
\usepackage{dcolumn}
\usepackage{bm}
\usepackage{siunitx}
\usepackage{color}
\usepackage[colorlinks,linkcolor=blue, anchorcolor=blue, urlcolor=blue, citecolor=blue]{hyperref}
\usepackage{multirow}
\newcommand{\ket}[1]{\mbox{$ | #1 \rangle $}}
\begin{document}
	
	
	\title{Experimental quantum key distribution secure against malicious devices}

	\author{Wei Li}
	\thanks{These authors contributed equally to this work.}	
	\affiliation{
	Hefei National Laboratory for Physical Sciences at the Microscale and Department of Modern Physics, University of Science and Technology of China, Hefei 230026, China}
	\affiliation{
	Shanghai Branch, CAS Center for Excellence in Quantum Information and Quantum Physics, University of Science and Technology of China, Shanghai 201315, China}
	\affiliation{
	Shanghai Research Center for Quantum Sciences, Shanghai 201315, China}

	\author{Víctor Zapatero}
	\thanks{These authors contributed equally to this work.}	\affiliation{Escuela de Ingeniería de Telecomunicaci$\acute{o}$n, Department of
		Signal Theory and Communications, University of Vigo, Vigo E-36310, Spain}
	\author{Hao Tan}
	\author{Kejin Wei}
	\author{Hao Min}
	\author{Wei-Yue Liu}
	\author{Xiao Jiang}
	\author{Sheng-Kai Liao}	
	\affiliation{
	Hefei National Laboratory for Physical Sciences at the Microscale and Department of Modern Physics, University of Science and Technology of China, Hefei 230026, China}
	\affiliation{
	Shanghai Branch, CAS Center for Excellence in Quantum Information and Quantum Physics, University of Science and Technology of China, Shanghai 201315, China}
	\affiliation{
	Shanghai Research Center for Quantum Sciences, Shanghai 201315, China}
	\author{Cheng-Zhi Peng}	
	\affiliation{
	Hefei National Laboratory for Physical Sciences at the Microscale and Department of Modern Physics, University of Science and Technology of China, Hefei 230026, China}
	\affiliation{
		Shanghai Branch, CAS Center for Excellence in Quantum Information and Quantum Physics, University of Science and Technology of China, Shanghai 201315, China}
	\affiliation{
		Shanghai Research Center for Quantum Sciences, Shanghai 201315, China}
	\affiliation{QuantumCTek Co., Ltd., Hefei, Anhui 230088, China}	
	\author{Marcos Curty}
	\affiliation{Escuela de Ingeniería de Telecomunicaci$\acute{o}$n, Department of
	Signal Theory and Communications, University of Vigo, Vigo E-36310, Spain}
	\author{Feihu Xu}
	\author{Jian-Wei Pan}
	\affiliation{
	Hefei National Laboratory for Physical Sciences at the Microscale and Department of Modern Physics, University of Science and Technology of China, Hefei 230026, China}
	\affiliation{
	Shanghai Branch, CAS Center for Excellence in Quantum Information and Quantum Physics, University of Science and Technology of China, Shanghai 201315, China}
	\affiliation{
	Shanghai Research Center for Quantum Sciences, Shanghai 201315, China}

	\begin{abstract}
		The fabrication of quantum key distribution (QKD) systems typically involves several parties, thus providing Eve with multiple opportunities to meddle with the devices. As a consequence, conventional hardware and/or software hacking attacks pose natural threats to the security of practical QKD. Fortunately, if the number of corrupted devices is limited, the security can be restored by using redundant apparatuses. Here, we report on the demonstration of a secure QKD setup with optical devices and classical post-processing units possibly controlled by an eavesdropper. We implement a 1.25 \si{\giga\hertz} chip-based measurement-device-independent QKD system secure against malicious devices on \emph{both} the measurement and the users' sides. The secret key rate reaches 137 bps over a 24 \si{\decibel} channel loss. Our setup, benefiting from high clock rate, miniaturized transmitters and a cost-effective structure, provides a promising solution for widespread applications requiring uncompromising communication security.
	\end{abstract}

	\maketitle

	Quantum key distribution (QKD)~\cite{BB84} enables the generation of information-theoretically secure cryptographic keys between two distant parties (Alice and Bob). Its security relies on fundamental properties of quantum mechanics together with certain assumptions~\cite{review1,review2}.~One crucial assumption is that Alice's and Bob's devices are honest, {\it i.e.}, they follow the prescriptions of the protocol and do not intentionally leak their internal information to the eavesdropper (Eve). In view of the many hardware and software Trojan horse attacks (THAs) against conventional cryptographic systems~\cite{Gligor,Zander,Yang,Becker} reported recently, this assumption seems however unjustified and over-optimistic. Indeed, the fabrication of QKD systems is a complex process that may involve several parties, which design, manufacture, package and distribute the different components. This provides Eve with multiple opportunities to meddle with the devices. We note that even device-independent QKD~\cite{Mayers,Scarani} is vulnerable to this type of attacks based on malicious devices~\cite{Barrett}.
	
	So far, the security of all implementations of QKD, e.g., decoy-state based BB84 schemes ~\cite{liao2017satellite,boaron2018secure}, measurement-device-independent (MDI) QKD~\cite{Comandar2016a,yin2016measurement,Wei2019a} and the recent twin-field QKD~\cite{liu2019experimental,zhong2019proof,Minder,Wang}, rely on the implicit assumption that the users' devices are honest. However, verifying the proper functioning of {\it all} the elements within a QKD setup is probably unfeasible in practice~\cite{Adee}. Note that even slight modifications of a few transistors in a chip might compromise the security. Also, Eve could circumvent any post-fabrication test by designing attack triggers based on an unlikely sequence of events~\cite{Yang,Becker}.
	
\begin{figure}[!htbp]
	\centering
	\includegraphics[width=0.6\linewidth]{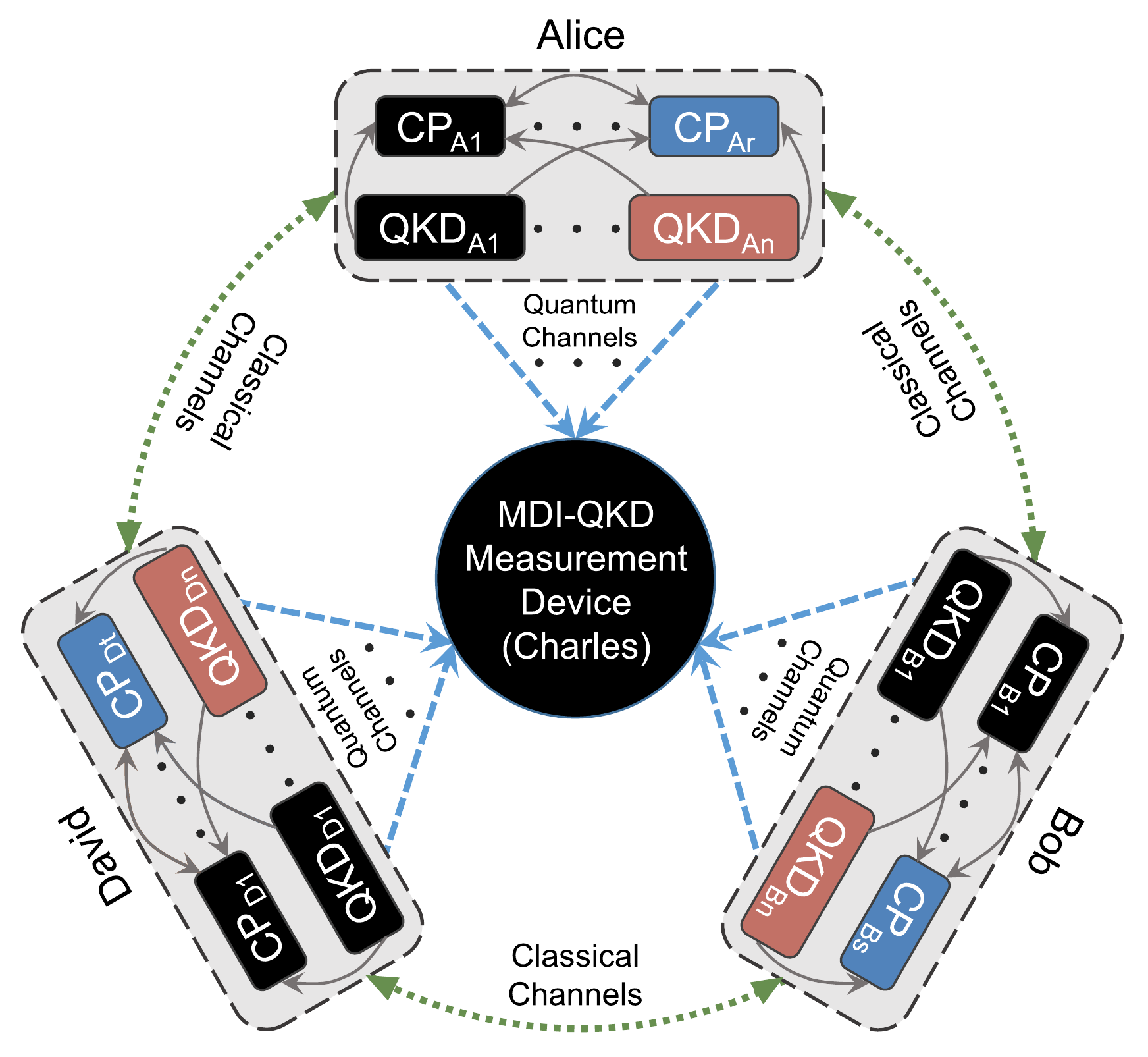}
	\caption{Diagrammatic representation of a MDI-QKD network with multiple QKD modules and classical post-processing (CP) units at each user. MDI-QKD allows the measurement module to be totally untrusted, which means that there is no need for redundant measurement devices. Meanwhile, the users possess various devices of each type, thus allowing for a restricted number of them to be corrupted. As an example, we mark in black a possible combination of malicious devices.}
	\label{fig:intro}
	\end{figure}

	A possible solution to overcome this problem and relax the security assumptions of QKD has been recently proposed in~\cite{Curty}. It uses \emph{redundant} devices together with secure multiparty computation techniques~\cite{Cramer,Maurer}, particularly, verifiable secret sharing (VSS). The proposal distinguishes two main types of devices: the ``QKD modules" (which mainly contain the optical/quantum components of the system) used to generate correlated raw data between Alice and Bob via quantum communication, and the ``classical post-processing (CP) units" that post-process this data to distill a secret key. By furnishing each of Alice and Bob with multiple QKD modules and CP units, it is possible to distribute secure keys even if some of these devices are malicious and controlled by Eve. More precisely, the requirement is that at least one pair of QKD modules ---that is, one at Alice's lab and one at Bob's lab connected to each other by a quantum channel--- is honest, and more than two thirds of Alice's (Bob's) CP units are honest too. Note that no security is possible with a fewer number of honest devices. Moreover, the combination of the approach in~\cite{Curty} with MDI-QKD~\cite{Curty1} enables a novel QKD network structure (see Fig.~\ref{fig:intro}), where the central relay may be fully untrusted and the users may have devices from corrupted vendors.

	Here, we demonstrate a QKD system secure against malicious devices. On the theoretical side, we put the ideas in~\cite{Curty} into practice by specifically devising a scheme with an improved multiparty post-processing procedure that requires a single privacy amplification step and minimises the authentication cost, thus making it more efficient and experimental-friendly. On the experimental side, we demonstrate polarization-encoding MDI-QKD with three integrated chip transmitters, and realize such multiparty post-processing procedure based on a VSS scheme. The chip-based MDI-QKD system is operated at a clock rate of 1.25 \si{\giga\hertz} which is among the highest reported so far. In so doing, we demonstrate secure QKD with a minimum of trust, {\it i.e.}, the setup is robust against malicious QKD modules, CP units, and measurement units connecting the QKD modules.

\begin{figure*}[!htbp]
	\centering
	\includegraphics[width=\textwidth]{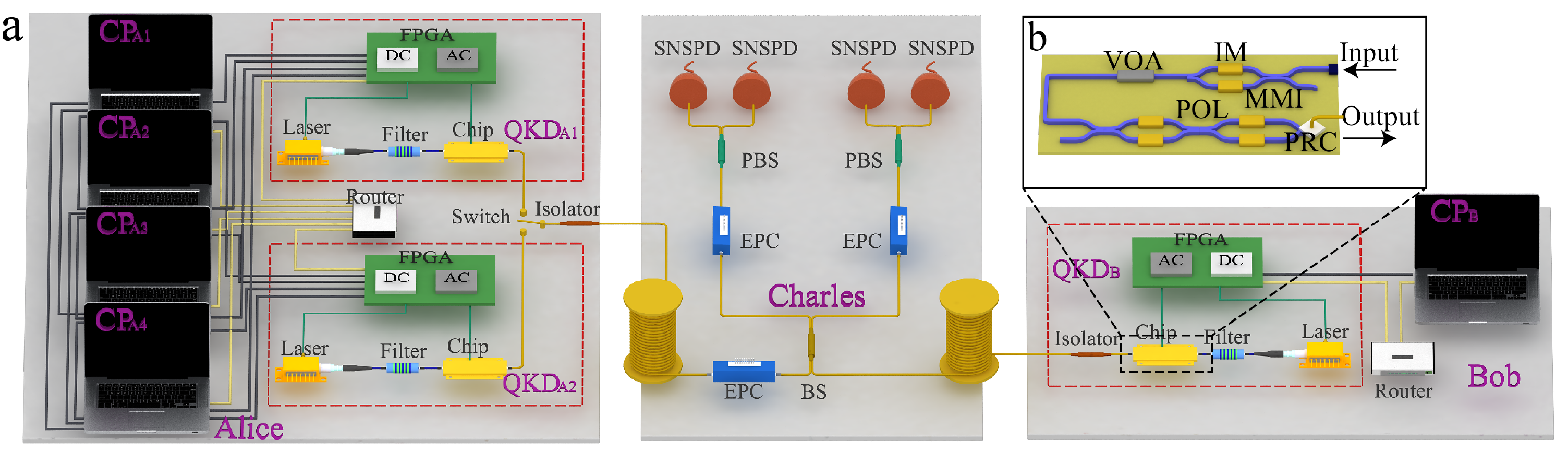}
	\caption{\textbf{a.} Depiction of the experimental MDI-QKD setup. Alice holds two QKD modules, $\{\textrm{QKD}_{\textrm{A}_{j}}\}_{j=1}^{2}$, and four CP units, $\{\textrm{CP}_{\textrm{A}_{l}}\}_{l=1}^{4}$, and, for simplicity, Bob has one QKD module, QKD$_{\rm B}$, and one CP unit, CP$_{\rm B}$, which are assumed to be trusted. This setup protects against one malicious device of each kind at Alice's side, as well as against a malicious Charles.~\textbf{b.} Schematic of the Si chip. It integrates an intensity modulator (IM), a polarization modulator (POL) and a variable optical attenuator (VOA). See the main text for the detailed description of the different elements.}
	\label{fig:setup}
	\end{figure*}

\textbf{Protocol.} A naive approach against malicious devices, also based on redundancy, is to simply take the XOR operation of the keys generated by different systems. By doing so, a single malicious system cannot guess the final key, unless all systems are malicious. However, such approach fails to guarantee the correctness of the final key and is suboptimal in the key rate. Ref.~\cite{Curty} proposes a method which guarantees both the secrecy and the correctness of the final key. The main idea is to use a VSS scheme~\cite{Cramer} to protect against malicious CP units, together with a dedicated privacy amplification technique to protect against malicious QKD modules. In particular, we consider a scenario (see Fig.~\ref{fig:setup}a), where Alice holds two QKD modules, $\{\textrm{QKD}_{\textrm{A}_{j}}\}_{j=1}^{2}$, and four CP units, $\{\textrm{CP}_{\textrm{A}_{l}}\}_{l=1}^{4}$. This configuration protects against one malicious device of each kind at Alice's side. On the other hand, for simplicity, we consider that Bob holds a single QKD module, $\textrm{QKD}_{\textrm{B}}$, and a single CP unit, $\textrm{CP}_{\textrm{B}}$, which are both trusted. We carefully allocate the post-processing tasks between Alice and Bob such that most of the required operations are actually conducted multipartily by Alice's four CP units. The considered protocol is described in detail in Section I of the Supplemental Material (SM)~\cite{supplemental}.
		
To defeat malicious CP units, VSS allows the raw key generated by a QKD module to be post-processed in a distributed way such that no CP unit can obtain any information about the final key. Also, the resulting secret key length is optimal in the sense that it matches the one attainable in the scenario with honest devices (disregarding the authentication cost). The post-processing tasks we implement multipartily include sifting, parameter estimation, error correction (EC), error verification (EV) and privacy amplification (PA). We note that the distributed post-processing can introduce extra authentication cost. In order to minimise this cost, we choose a low-density parity-check (LDPC) code~\cite{Luby1998a,Chen2005} to reconcile Bob's key with Alice's key and deliberately devise a protocol with minimum authenticated communication.
	
On the other hand, PA~\cite{PA} is applied to remove any information that could be revealed to Eve due to the presence of malicious QKD modules. We use a single PA step to compress the concatenation of the sifted keys obtained from the raw data provided by $\{\textrm{QKD}_{\textrm{A}_{j}}\}_{j=1}^{2}$, to get a final secret key. Toeplitz matrices acting as two-universal hash function are selected by a random bit string (RBS) generation protocol~\cite{supplemental}, which is also performed by all four $\textrm{CP}_{\textrm{A}_{l}}$ distributedly.	

In the presence of malicious devices, the finite-key secret length of, say, the $j$-th QKD pair, can be written as~\cite{supplemental}
\begin{equation}\label{l_j}
l_{j}=\biggl\lfloor{S_{11,\rm Z}^{j,\rm L}\left[1-h(\phi_{11,\rm Z}^{j,\rm U})\right]-\lambda_{\rm EC}^{j}-t_{\rm EV}-\log_{2}\left(\frac{1}{4\epsilon_{\rm PA}^{2}\delta}\right)}\biggr\rfloor
	\end{equation}
where $S_{11,\rm Z}^{j,\rm L}$ ($\phi_{1,\rm Z}^{j,\rm U}$) is a lower (upper) bound on the number of single-photon contributions (single-photon phase-error rate) in the sifted keys, $h(\cdot)$ is the binary entropy function, $\lambda_{\mathrm{EC}}^{j}$ is an upper bound on the number of bits revealed by EC, $t_{\rm EV}$ is the size of the EV tag, $\epsilon_{\rm PA}$ is the error probability of the PA step and $\delta\in(0,1)$.

The parameter estimation step builds on the finite decoy-state bounds derived in~\cite{Curty2} and is redundantly performed by three $\textrm{CP}_{\textrm{A}_{l}}$ to assure the presence of a majority of honest units in the process. The reader is referred to Section II of the SM~\cite{supplemental} for more details on the parameter estimation and the finite-key analysis.

\textbf{\textbf{Setup.}} The experimental setup is illustrated in Fig.~\ref{fig:setup}a. Each QKD module consists of an off-chip distributed feedback (DFB) laser, together with a silicon photonic chip (see Fig.~\ref{fig:setup}b). For $j=1,2$, the $j$-th pair of QKD modules $(\textrm{QKD}_{\textrm{A}_{j}},\textrm{QKD}_{\textrm{B}})$ implements a chip-based polarization-encoding MDI-QKD link. An optical switch at Alice's station allows to select the operating QKD module, \textit{i.e.}, $\textrm{QKD}_{\textrm{A}_{1}}$ or $\textrm{QKD}_{\textrm{A}_{2}}$. Pulses from Alice and Bob interfere at a linear-optics Bell state measurement (BSM) device within Charles' node.
	
	Alice (Bob) independently gain-switches a 1550~\si{\nano\meter} DFB laser diode to generate 1.25 \si{\giga\hertz} phase-randomized weak coherent pulses (PR-WCPs) with a full width at half maximum (FWHM) of 163~\si{\pico\second}. A Hong-Ou-Mandel interference visibility of 47.4\% is achieved at Charles' node without using the laser injection technique~\cite{Comandar2016a}. This benefits from a sharp flat-top narrowband (5 \si{\giga\hertz}) tunable filter. The center wavelength and bandwidth of the filter have been selected carefully in order to achieve a high interference visibility, which is crucial to obtain a low quantum bit error rate (QBER) in MDI-QKD. This greatly reduces the complexity of the source and may benefit the integration of the laser diode onto the chip~\cite{Semenenko2019,Agnesi2019}.
	
	Each transmitter chip (Fig.~\ref{fig:setup}b) is fabricated in a standard Si photonic foundry process offered by IMEC. It integrates a Mach-Zehnder interferometer (MZI) acting as an intensity modulator (IM), a three-stage p-i-n diode acting as a variable optical attenuator (VOA), and a polarization rotator combiner (PRC) combined with an additional stage of MZI operating as a polarization modulator (POL). In particular, the PRC is realized by a chip-to-fiber 2D grating coupler and the active components of the IM and POL include a thermo-optic modulator, which provides the bias, and an electro-optic modulator, which has a modulation bandwidth of 21 \si{\giga\hertz}. By harnessing multimode interferometers (MMI), the MZI-type IM demonstrates a dynamic extinction ratio (ER) of 19~\si{\decibel}. Such ER allows the lowest and the largest intensities to differ in about two orders of magnitude, which suffices for a tight decoy-state parameter estimation. The VOA attenuates the generated light to single-photon level, and it is laid before the POL to prevent polarization-dependent loss. Also, to prevent optical THAs~\cite{Gisin,lucamarini2015practical,Wang2020}, an off-chip isolator is added too. The POL has a dynamic polarization ER of 20 \si{\decibel}; this guarantees a high-fidelity preparation of the four BB84 polarization states. The generation of such states at a \si{\giga\hertz} clock rate is a quite challenging experimental task. For this purpose, a field-programmable gate array (FPGA) board with an analog circuit (AC) and a digital circuit (DC) is designed~\cite{Liu2020}, featuring 7.5~\si{\volt}$ _{\mathrm{pp}} $ and low modulation noise.
	
	Charles performs a BSM using a 50:50 beam-splitter (BS), two polarization beam-splitters (PBS) and four superconducting nanowire single-photon detectors (SNSPD). According to our measurement, the total insertion loss of the BSM is 1.1~\si{\decibel} and the quantum efficiency of the SNSPD is 49.5\%. A successful measurement event occurs when two SNSPDs associated to orthogonal polarizations are triggered. To achieve good interference, the photons must have the same polarization, timing, and spectrum modes. Three electrical polarization controllers (EPC) are used to align Alice's and Bob's polarization reference frames with a precision up to 28 \si{\decibel}. Regarding the timing mode, a clock board distributes synchronized clock signals among all stations with a root-mean-square jitter of 8 \si{\pico\second}. The FPGA DC module can adjust the delay in steps of~1.5 \si{\pico\second}, which is accurate enough compared to the PR-WCP FWHM. Thirdly, the spectra of the laser diodes are matched by tuning the driving current and the temperature. The temperature is stabilised by a proportional–integral–derivative loop with an instability of 3 \si{\milli\kelvin}. Considering the temperature coefficient of the DFB laser (0.09 \si{\nano\meter}/\si{\kelvin}), a good spectrum overlap is maintained throughout the experiment.

\begin{figure}[!htbp]
		\centering
		\includegraphics[width=0.7\linewidth]{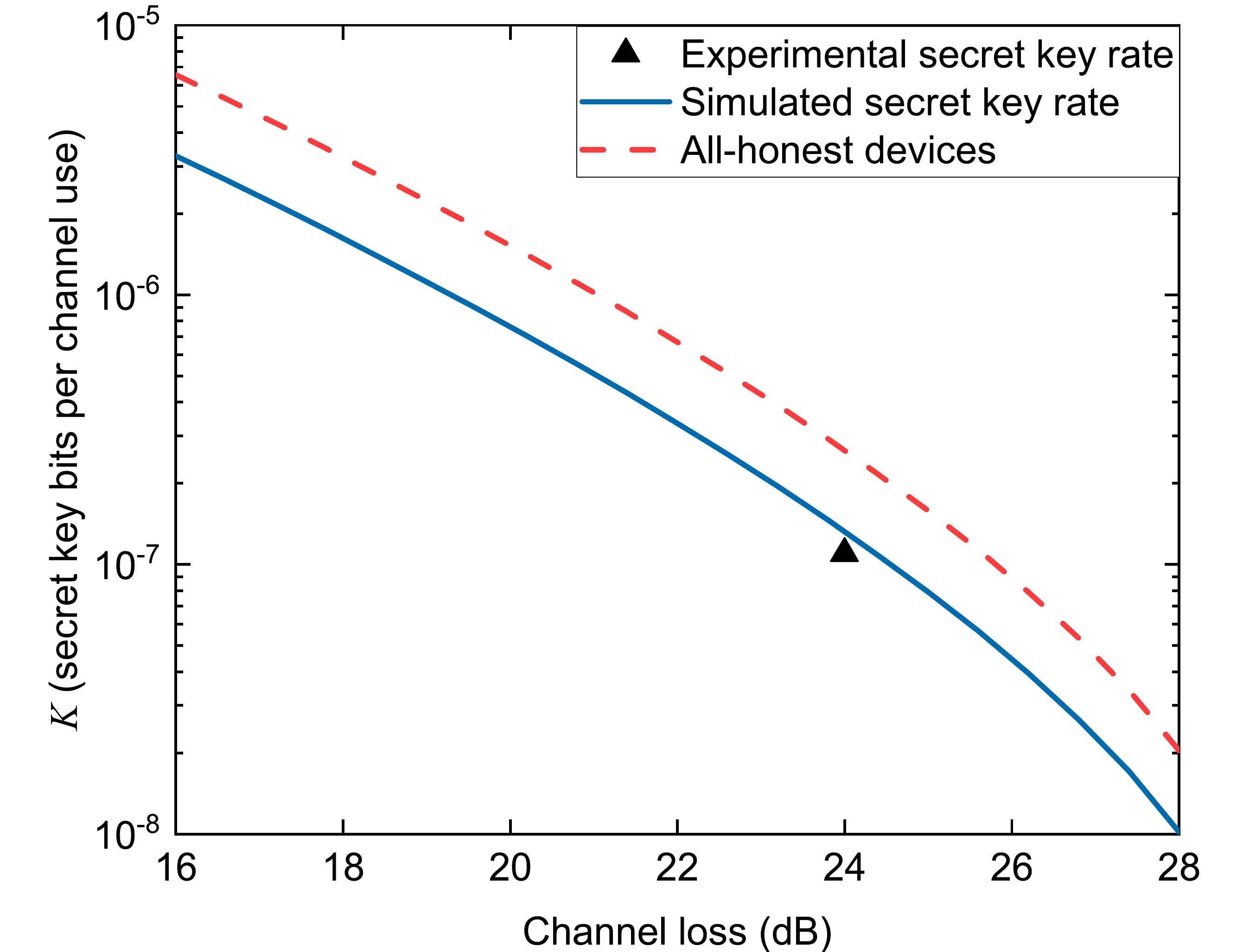}
		\caption{Simulation (lines) and experimental result (solid black triangle) of the secret key rate as a function of the channel loss. The achieved experimental secret key rate for a total emulated channel loss of 24 dB is $ 1.1\times 10^{-7} $ bits/pulse. The solid line corresponds to a simulation of the secret key rate in the setting with dishonest devices described by the experiment. The dashed line corresponds to the standard setting where the parties assume that all their devices are honest. The secret key rate in this latter scenario roughly doubles the one with dishonest devices. This is so because of the need to discard the key generated by the potentially malicious QKD pair via PA when dishonest devices are considered. Moreover, this approximate factor two shows that the extra authentication cost required in the experiment due to the use of redundant devices is negligible.}
		\label{fig:rate}
	\end{figure}

\textbf{Results.} In the experiment, the quantum communication runs over an emulated attenuation of 12 \si{\decibel}---corresponding to about 60 \si{\kilo\meter} of standard fiber---on each side. For both QKD pairs, $(\mathrm{QKD}_{\mathrm{A}_{j}},\mathrm{QKD}_{\mathrm{B}})$ with $j=1,2$,
the four intensities MDI-QKD protocol~\cite{X.B.Wang} is implemented, where Alice and Bob use a single signal intensity in the Z basis (devoted to key generation) and three decoy intensities in the X basis (devoted to parameter estimation). Fig.~\ref{fig:rate} shows the achieved experimental secret key rate together with a simulation of the secret key rate as a function of the channel loss for a channel model matching the experimental parameters. These parameters are shown in Section III of the SM~\cite{supplemental}.

We run the experiment for 16000 seconds per QKD pair (or, equivalently, we send a total number of $N=2\times10^{13}$ pulses per QKD module in each pair). Overall, we observe a typical QBER of 2.1\% (27.9\%) in the Z (X) basis and extract a total secret key of 4386592 bits. The secret key rate is $ 1.1\times 10^{-7} $ bits per pulse or 137 bits per second. Notably, in Fig.~\ref{fig:rate}, we show that the cost of the extra classical authentication is negligible compared to the total secret key length. Therefore, the resulting secret key rate is approximately half of what could have been obtained with the same amount of quantum communication but assuming all the devices to be honest. This feature comes from the need to discard the key generated by the potentially malicious QKD pair via PA.

\begin{figure}[!htbp]
		\centering
		\includegraphics[width=\linewidth]{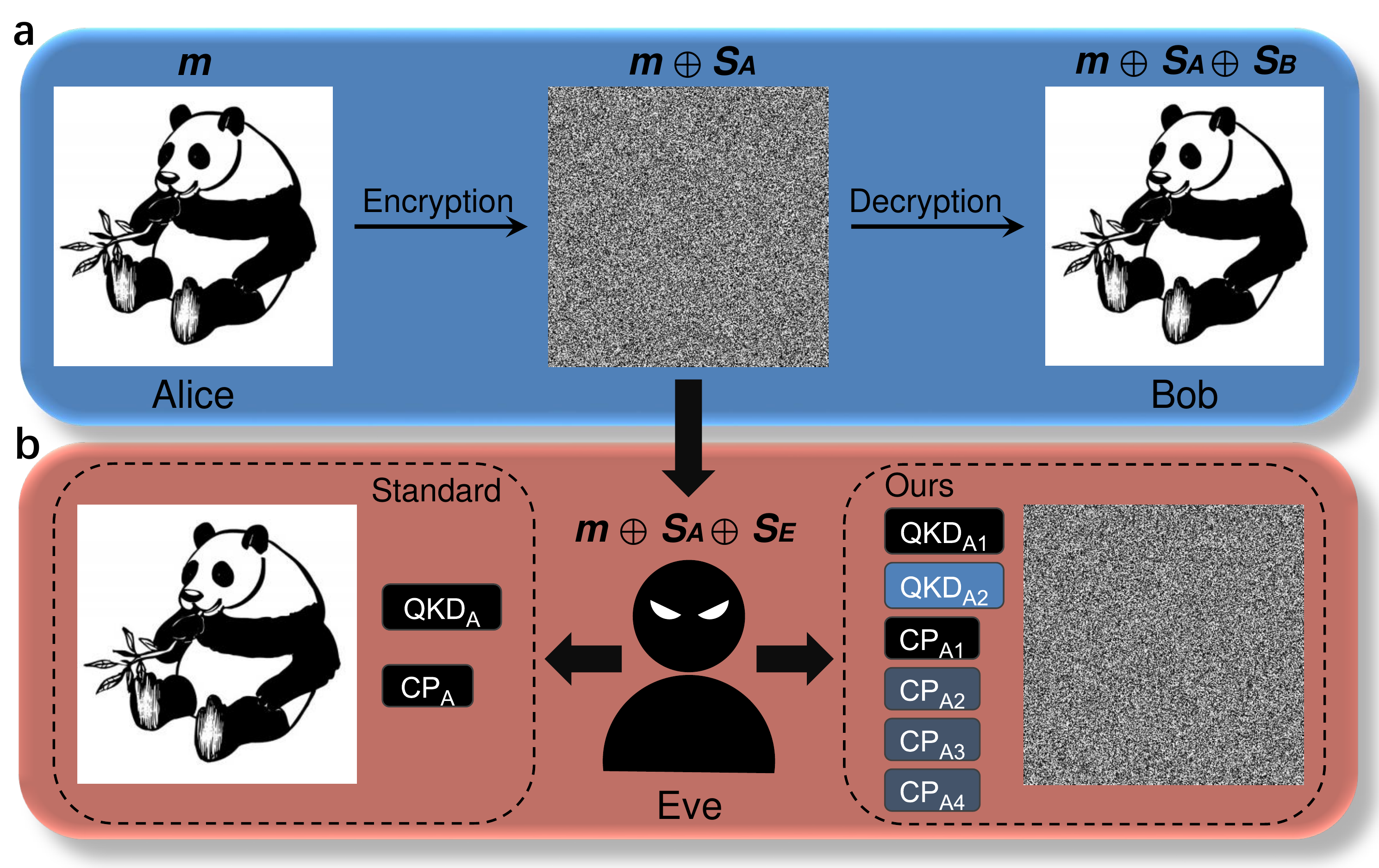}
		\caption{\textbf{a.} Encryption and decryption with the shared secret key. Alice encrypts a $ 300\times300 $ pixels grayscale image of a panda, $m$, with her final key, $S_{\rm A}$, and the one-time pad encryption scheme. Bob decrypts the image correctly using his final key $S_{\rm B}=S_{\rm A}$.~\textbf{b.} Illustration of Eve's failed attempt to decrypt the encrypted picture. Despite Eve's significant knowledge about the sifted key, $Z_{\rm A}$, due to both her possible intervention in the quantum channel and the information revealed by the corrupted devices ---assumed to be QKD$_{\mathrm{A}_{1}}$ and CP$_{\mathrm{A}_{1}}$ for illustration purposes here---, her attempt to decrypt the intercepted picture using $S_{\rm E}$ (defined in the main text) yields a white noise output, fully uncorrelated to the actual picture. However, if, as in standard QKD, only one QKD module and one CP unit are used, Eve's attempt is successful unless both of them are honest.}
		\label{fig:decrypt}
	\end{figure}

	To further illustrate our results, we consider that Alice uses her final key $S_{\rm A}$ to encrypt a message $m$ with the one-time pad scheme and sends the encrypted message to Bob. To reconstruct $S_{\rm A}$, Alice herself collects all the shares held by her CP units and then applies majority voting followed by an XOR operation. Also, we assume that, say, QKD$_{\mathrm{A}_{1}}$ and CP$_{\mathrm{A}_{1}}$ are dishonest. In particular, this implies that the sifted key of the first pair, $Z_{\rm A}^{1}$, and the second, third and fourth shares of $Z_{\rm A}^{2}$ could be known to Eve. Since, in addition, the privacy amplification function $h_{\rm PA}$ is made public, Eve could apply $h_{\rm PA}$ on the concatenation $[Z_{\rm A}^{1},\bigoplus_{k=2}^{4}Z_{\mathrm{A},k}^{2}]$ (where $Z_{\mathrm{A},k}^{2}$ stands for the $k$-th share of $Z_{\rm A}^{2}$), which compresses all the information about $Z_{\rm A}$ held by the malicious devices. However, it can be shown that the resulting string, say $S_{\rm E}=h_{\rm PA}([Z_{\rm A}^{1},\bigoplus_{k=2}^{4}Z_{\mathrm{A},k}^{2}])$, is totally uncorrelated to Alice's final key $S_{\rm A}=h_{\rm PA}(Z_{\rm A})$. As a consequence, the outcome of Eve's decryption attempt $S_{\rm E}\oplus{}S_{\rm A}\oplus{m}$ is a fully random string totally uncorrelated to $m$. Fig.~\ref{fig:decrypt} exemplifies this decorrelation by using a picture as the plain message $m$ and showing that Eve's attempt to decrypt the figure with $S_{\rm E}$ yields a white noise picture. Of course, a similar conclusion follows for any other combination of corrupted devices, as long as a single QKD$_{\mathrm{A}_{j}}$ (CP$_{\mathrm{A}_{l}}$) is corrupted at most.
	
\textbf{\textbf{Discussion.}} Our experiment demonstrates the feasibility of secure QKD in the presence of a restricted number of optical devices and post-processing units possibly corrupted and controlled by Eve. The QKD protocol we have implemented is a chip-based MDI-QKD scheme operating at a high clock rate. Following the proposal in~\cite{Curty}, the use of redundant QKD modules and CP units allows us to combine VSS techniques~\cite{Maurer} with PA~\cite{PA} in order to establish the security of the final keys. To be precise, the experiment shows that the typical post-processing steps standard in QKD can be implemented multipartily in a distributed way if necessary. Although, for simplicity, our scheme only protects against one corrupted QKD module (CP unit) at Alice's side, it could easily be extended to protect against more corrupted devices both at Alice's and Bob's labs. Also, the only essentially different task implemented at Bob's side in our experiment, that is to say, the decoding algorithm of EC, can be executed multipartily too. For completeness, we have also demonstrated this alternative using Alice's CP units, and a detailed explanation is given in the SM. The implemented protocol can also be combined with a twin-field QKD protocol~\cite{lucamarini2018overcoming} to enhance both the security and the distance.
	
	The trade-off we pay is in terms of secret key rate and equipment, although it is not critical in either case. Indeed, in order to provide the highest level of security ---as promised by QKD theorists---, such sacrifices are probably more than justified by the increasing frequency and variety of security breaches affecting conventional communication systems. In this scenario, the reported experiment is a solid step forward towards foiling an unpreventable loophole in the security of QKD.

\textbf{Acknowledgements.} This work was supported by the National Key Research and Development (R\&D) Plan of China (under Grants No. 2018YFB0504300 and 2017YFA0304000), the National Natural Science Foundation of China (under Grants No. 61771443 and No. 61705048), the Anhui Initiative in Quantum Information Technologies, the Shanghai Municipal Science and Technology Major Project (Grant No.2019SHZDZX01) and the Chinese Academy of Sciences. V.Z. gratefully acknowledges support from a FPU scholarship from the Spanish Ministry of Education. M.C. acknowledges support from the Spanish Ministry of Economy and Competitiveness (MINECO), the Fondo Europeo de Desarrollo Regional (FEDER) through grant TEC2017-88243-R, and the European Union's Horizon 2020 research and innovation programme under the Marie Sklodowska-Curie grant agreement No 675662 (project QCALL).

\clearpage

\textbf{{\large SUPPLEMENTAL MATERIAL}}
\vspace{0.5cm}

\section{Protocol description} 
We follow the efficient MDI-QKD scheme proposed in~\cite{X.B.Wang_}. In this scheme, Alice and Bob use a single signal intensity for the basis Z, devoted to key extraction, and perform parameter estimation with the basis X alone, for which they use three different signal intensities.\\

\begin{figure}[!htbp]
	\centering 
	\includegraphics[width=11cm,height=5cm]{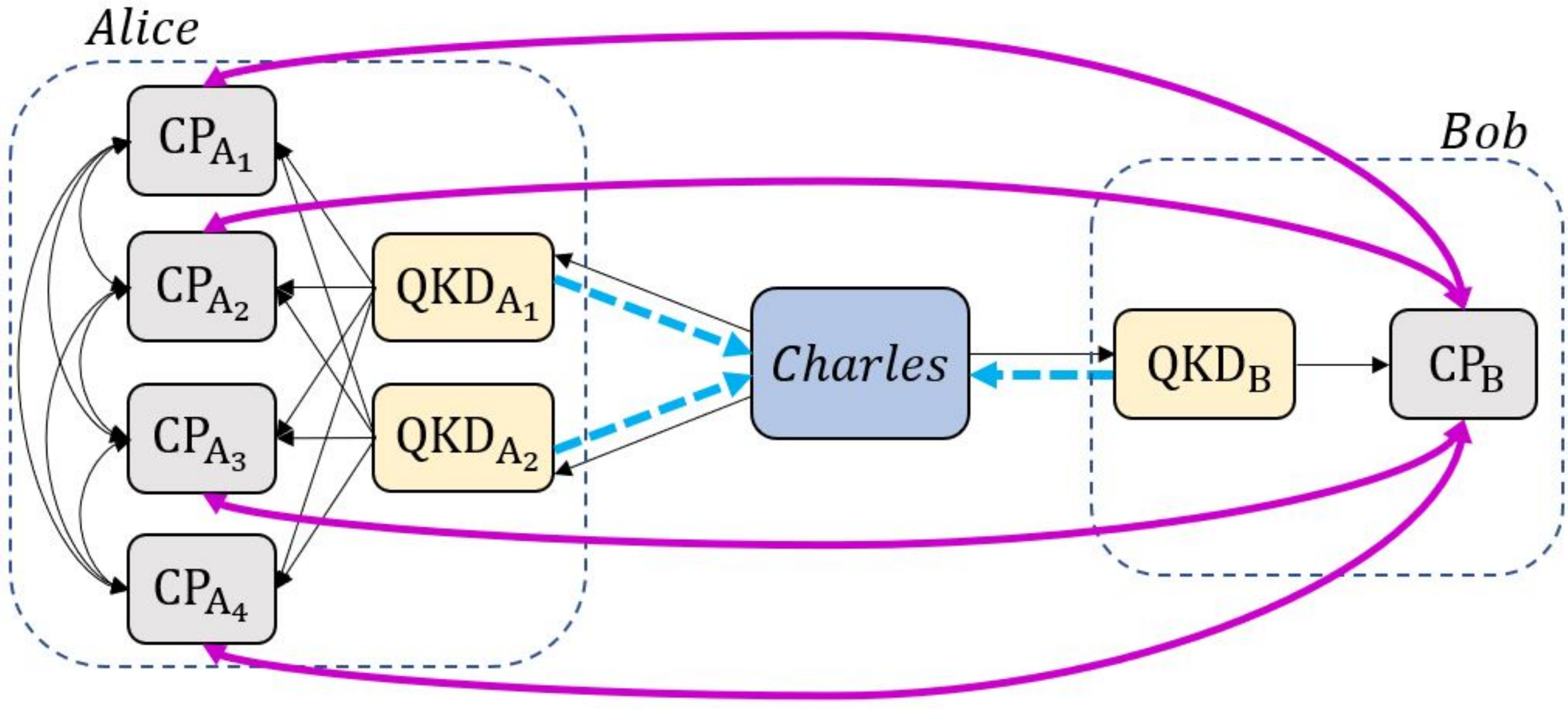}\\ 
	\caption{Schematic of the setup implementing the protocol presented in the main text. While Alice holds two QKD modules, $\{\textrm{QKD}_{\textrm{A}_{j}}\}_{j=1}^{2}$, and four CP units, $\{\textrm{CP}_{\textrm{A}_{l}}\}_{l=1}^{4}$, Bob holds a single module, $\textrm{QKD}_{\textrm{B}}$, and a single unit, $\textrm{CP}_{\textrm{B}}$, for simplicity.
		Each pair of modules, $(\textrm{QKD}_{\textrm{A}_{j}},\textrm{QKD}_{\textrm{B}})$ with $j=1,2$, is used to implement a MDI-QKD link using Charles' central node. The two pairs of keys generated by these links are post-processed by the CP units. Precisely, on Alice's side, the post-processing is performed in a multiparty setting using all four units $\{\textrm{CP}_{\textrm{A}_{l}}\}_{l=1}^{4}$, while, for simplicity, a standard single-party post-processing is applied on Bob's side. In this scenario, the protocol presented in the main text is secure if one of Alice's QKD modules and/or one of Alice's CP units are corrupted at most. The quantum channels are marked with dashed blue arrows, while every other arrow in the figure is solid and represents a classical channel. All of Alice's units are connected to each other by links which are assumed to be physically shielded and to connect only the desired units. Thus, further encryption and authentication is not required for them. The same applies to the links that connect a QKD module to a CP unit inside each lab, through which the raw key material and some protocol information is sent to the units for the post-processing. The classical links connecting a $\textrm{CP}_{\textrm{A}_{l}}$ with $\textrm{CP}_{\textrm{B}}$ are authenticated (solid purple arrows). Finally, the central node (Charles) is untrusted and thus no assumptions are made about the classical links between this node and the QKD modules.}
	\label{fig:setup2}
\end{figure}

We shall use the nomenclature of ``QKD modules" and ``CP units" presented in the main text and originally introduced in~\cite{Curty_}. In the experiment, Alice holds two QKD modules, $\{\textrm{QKD}_{\textrm{A}_{j}}\}_{j=1}^{2}$, and four CP units, $\{\textrm{CP}_{\textrm{A}_{l}}\}_{l=1}^{4}$, and we assume that one device of each kind might be corrupted at most. For simplicity, Bob holds a QKD module, $\textrm{QKD}_{\textrm{B}}$, and a CP unit, $\textrm{CP}_{\textrm{B}}$, and we assume that both devices are trusted. In this scenario, two combinations of ``QKD pairs" are used in the protocol: $(\textrm{QKD}_{\textrm{A}_{1}},\textrm{QKD}_{\textrm{B}})$ and $(\textrm{QKD}_{\textrm{A}_{2}},\textrm{QKD}_{\textrm{B}})$. A QKD pair is corrupted if one of its modules is, which means that one of the previous QKD pairs might be corrupted at most (since at most one of Alice's QKD modules might be corrupted). A schematic of the setup is shown in Fig.~\ref{fig:setup2}.

In what follows, we describe the implemented QKD protocol step by step, from the quantum communication to the classical post-processing. For simplicity, the description assumes that the corrupted devices do not deviate from the protocol prescriptions. Nevertheless, as explained in the main text, the use of VSS~\cite{Maurer_} and privacy amplification techniques guarantee the security of the protocol against misbehaving corrupted devices. This said, the protocol runs as follows.

For $j=1,2$, $\textrm{QKD}_{\textrm{A}_{j}}$ and $\textrm{QKD}_{\textrm{B}}$ create the pairs of strings $(r^{j},a^{j})$ and $({r'}^{j},b^{j})$, respectively. While $r^{j}$ and ${r'}^{j}\in\left\{0,1\right\}^{N}$ are fully random, $a^{j}$ and $b^{j}\in\left\{\lambda,\mu,\nu,\omega\right\}^{N}$ verify $P\bigl[a_{i}^{j}=\lambda\bigr]=P\bigl[b_{i}^{j}=\lambda\bigr]=q_{\rm Z}$ and $P\bigl[a_{i}^{j}=a\bigr]=P\bigl[b_{i}^{j}=a\bigr]=(1-q_{\rm Z})p_{a}$, for $a\in{}\{\mu,\nu,\omega\}$ and $i=1,2,...,N$. For convenience, we define ${A}=\{\mu,\nu,\omega\}$. On each side, the intensity $\lambda$ determines the use of the basis Z, and the basis X is used otherwise.\\

Let us now focus on a single QKD pair, say, the $j$-th one ($j=1,2$). For $i$ ranging from $1$ to a pre-specified number of rounds, $N$, steps 1 to 6 are repeated.
\begin{enumerate}
	\item\textbf{State preparation.} $\textrm{QKD}_{\textrm{A}_{j}}$ ($\textrm{QKD}_{\textrm{B}}$) prepares a phase-randomised weak coherent pulse (PR-WCP) with intensity $a_{i}^{j}$ ($b_{i}^{j}$) in the BB84 state defined by both $a_{i}^{j}$ and the bit value $r_{i}^{j}$ ($b_{i}^{j}$ and the bit value ${r'}_{i}^{j}$).
	\item\textbf{Transmission.} $\textrm{QKD}_{\textrm{A}_{j}}$ and $\textrm{QKD}_{\textrm{B}}$ send the states to Charles via the quantum channel. 
	\item\textbf{Measurement.} If Charles is honest, he measures the received signals with a Bell state measurement (BSM). In any case, he sends the list of indexes of the successful BSM rounds, $ \mathcal{I}_j $ and the state information $ s_j $ to both modules. If he is honest, $s_{j}^{i}=0$ ($s_{j}^{i}=1$) if a successful BSM associated to the state $\ket{\psi^{-}}$ ($\ket{\psi^{+}}$) occurred. Note that in the experiment we use a standard linear optics BSM, and therefore, only $\ket{\psi^{-}}$ and $\ket{\psi^{+}}$ can be post-selected.
	\item\textbf{Distribution of data.} $\textrm{QKD}_{\textrm{B}}$ communicates $b^{j}|_{\mathcal{I}_j}$ and ${r'}^{j}|_{\mathcal{I}_j}$ (\textit{i.e.,} the restrictions of $b^{j}$ and ${r'}^{j}$ to the set of rounds indexed by $\mathcal{I}_j$) to $\textrm{CP}_{\rm B}$. Let $a^{j}|_{\mathcal{I}_j}$ denote the restriction of $a^{j}$ to the non-zero entries of $\mathcal{I}_j$ too, and let ${r}^{j}|_{\mathcal{I}_j,\rm X}$ (${r}^{j}|_{\mathcal{I}_j,\rm Z}$) be the restriction of ${r}^{j}$ to the set of rounds indexed by $\mathcal{I}_j$ where $\textrm{QKD}_{\textrm{A}_{j}}$ used basis X (Z). $\textrm{QKD}_{\textrm{A}_{j}}$ communicates $s_{j}$, $a^{j}|_{\mathcal{I}_j}$ and ${r}^{j}|_{\mathcal{I}_j,\rm X}$ directly to every $\textrm{CP}_{\textrm{A}_{l}}$, but uses the Share protocol of a VSS scheme to distribute shares of ${r}^{j}|_{\mathcal{I}_j,\rm Z}$ among them (see the post-processing section below). All $\textrm{CP}_{\textrm{A}_{l}}$ pairwise check the consistency of their copies of $s_{j}$, $a^{j}|_{\mathcal{I}_j}$ and ${r}^{j}|_{\mathcal{I}_j,\rm X}$ through authenticated channels. If a $\textrm{CP}_{\textrm{A}_{l}}$ finds an inconsistency, it raises a complaint through a (possibly simulated) broadcast channel and $\textrm{QKD}_{\textrm{A}_{j}}$ broadcasts $s_{j}$, $a^{j}|_{\mathcal{I}_j}$ and ${r}^{j}|_{\mathcal{I}_j,\rm X}$ to all the $\textrm{CP}_{\textrm{A}_{l}}$.
	\item\textbf{Sifting.} $\textrm{CP}_{\rm B}$ sends $b^{j}|_{\mathcal{I}_j}$ and ${r'}^{j}|_{\mathcal{I}_j,\rm X}$ (defined identically as ${r}^{j}|_{\mathcal{I}_j,\rm X}$ but with respect to ${r'}^{j}$) to the first three units of Alice, $\{\textrm{CP}_{\textrm{A}_{l}}\}_{l=1}^{3}$, through authenticated channels. We refer to this particular set of three CP units at Alice's side as $\mathcal{A}$ in what follows. Every unit in $\mathcal{A}$ builds a string of coincidences, $ \mathcal{Z}_{j}$, such that the $k$-th bit $ \mathcal{Z}_{j}^{k}=1 $ if $ a^{j}_{k}=b^{j}_{k}=\lambda $, and $ \mathcal{Z}_{j}^{k}=0 $ otherwise, where $k$ ranges the set of rounds indexed by $\mathcal{I}_j$. Similarly, every unit in $\mathcal{A}$ builds the strings $\mathcal{X}_{j}^{a,b}$ such that $ \mathcal{X}_{j}^{a,b,k}=1 $ if $ a^{j}_{k}=a,b^{j}_{k}=b $ (with $a,b\in\mathcal{A}$), and $ \mathcal{X}_{j}^{a,b,k}=0 $ otherwise. Each unit in $\mathcal{A}$ discards all the zero-entry data in order to sift:
	\begin{enumerate}
		\item{its shares of ${r}^{j}|_{\mathcal{I}_j,\textrm{Z}}$ into shares of the sifted key $Z_{\rm A}^{j}={r}^{j}|_{\mathcal{Z}_{j}}$},
		\item{the raw data ${r}^{j}|_{\mathcal{I}_j,\textrm{X}}$ (${r'}^{j}|_{\mathcal{I}_j,\textrm{X}}$) into the set of strings $\left\{{r}^{j}|_{\mathcal{X}_{j}^{a,b}}\right\}_{a,b\in{A}}$ $\left(\left\{{r'}^{j}|_{\mathcal{X}_{j}^{a,b}}\right\}_{a,b\in{A}}\right)$ used for parameter estimation.}
	\end{enumerate}
	Note that, by definition of the Share protocol, all three units in $\mathcal{A}$ hold the fourth share of $Z_{\rm A}^{j}$. Each of them applies the adequate bit flips on its copy of this share to properly correlate the sifted key $Z_{\rm A}^{j}$ with ${r'}^{j}|_{\mathcal{Z}_{j}}$, according to $s_{j}$ (see~\cite{Curty1_,Curty2_}). Identically, each of them performs the adequate bit flips on the parameter estimation strings ${r}^{j}|_{\mathcal{X}_{j}^{a,b}}$ ($a,b\in{A}$) to properly correlate them with the corresponding strings of Bob, ${r'}^{j}|_{\mathcal{X}_{j}^{a,b}}$, according to $s_{j}$.
	\item\textbf{Parameter estimation.}
	For every $a,b\in{A}$, each unit in $\mathcal{A}$ computes the numbers of errors
	\begin{equation}\label{error_rates}
	e_{j}^{a,b}=\sum_{k=1}^{\bigl|\mathcal{X}_{j}^{a,b}\bigr|}r_{k}^{j}\bigr|_{\mathcal{X}_{j}^{a,b}}\oplus{{r'}_{k}^{j}\bigr|_{\mathcal{X}_{j}^{a,b}}}.
	\end{equation}
	Here, $r_{k}^{j}\bigr|_{\mathcal{X}_{j}^{a,b}}$ (${r'}_{k}^{j}\bigr|_{\mathcal{X}_{j}^{a,b}}$) denotes the $k$-th bit of the corresponding string. Finally, using $\left|\mathcal{Z}_{j}\right|$, $\bigl|\mathcal{X}_{j}^{a,b}\bigr|$ and $e_{j}^{a,b}$, each unit in $\mathcal{A}$ computes a lower bound, $S_{11,\rm Z}^{j,\rm L}$, on the number of single-photon successes in $\mathcal{Z}_{j}$ and an upper bound, $\phi_{11,\rm Z}^{j,\rm U}$, on the single-photon phase-error rate associated to the single-photon successes in $\mathcal{Z}_{j}$. Then, each unit in $\mathcal{A}$ checks if~\cite{Victor}
	
	\begin{equation}\label{l_j_}
	l_{j}=\biggl\lfloor{S_{11,\rm Z}^{j,\rm L}\left[1-h(\phi_{11,\rm Z}^{j,\rm U})\right]-\lambda_{\rm EC}^{j}-t_{\rm EV}-\log_{2}\left(\frac{1}{4\epsilon_{\rm PA}^{2}\delta}\right)}\biggr\rfloor
	\end{equation}
	is greater than zero, where $h(\cdot)$ is the binary entropy function, $\lambda_{\rm EC}^{j}$ is an upper bound on the number of bits revealed by error correction (EC), $t_{\rm EV}=64$ bits is the size of the error verification (EV) tag, $\epsilon_{\rm PA}$ is the error probability of the privacy amplification, and $\delta\in(0,1)$ (see the next section of this Supplemental Material for further details). If $l_{j}\leq{0}$, each unit in $\mathcal{A}$ complains and $\textrm{QKD}_{\textrm{A}_{j}}$ aborts the protocol. Otherwise, each unit in $\mathcal{A}$ communicates ${\mathcal{Z}_{j}}$ and $l_{j}$ to $\textrm{CP}_{\textrm{A}_{4}}$, which applies majority voting (MV) on both items and constructs its shares of the sifted key $Z_{\rm A}^{j}$ too. Similarly, each unit in $\mathcal{A}$ sends ${\mathcal{Z}_{j}}$ to $\textrm{CP}_{\textrm{B}}$. $\textrm{CP}_{\textrm{B}}$ applies MV to decide on a single copy of ${\mathcal{Z}_{j}}$ and uses this copy to sift its raw string ${r'}^{j}|_{\mathcal{I}_j,\textrm{Z}}$ into the sifted key $Z_{\rm B}^{j}={r'}^{j}|_{\mathcal{Z}_{j}}$.\\
	
	Once steps 1 to 6 are performed for $j=1,2$ (and if the protocol does not abort), all $\textrm{CP}_{\textrm{A}_{l}}$ hold shares of the concatenated string $\textrm{Z}_{\rm A}=\textrm{Z}_{\rm A}^{1}\textrm{Z}_{\rm A}^{2}$, while $\textrm{CP}_{\textrm{B}}$ holds the concatenation $\textrm{Z}_{\rm B}=\textrm{Z}_{\rm B}^{1}\textrm{Z}_{\rm B}^{2}$.
	\item\textbf{Random permutation.}
	In order to prevent burst errors that could cause the failure of the information reconciliation (IR) step, each unit randomly permutes the bytes of the sifted key shares using a permutation matrix known a priori to all of them. This matrix can be reused in arbitrarily many sessions.
	\item\textbf{Information reconciliation.}
	Each $\textrm{CP}_{\textrm{A}_{l}}$ constructs its shares of the syndrome $sy_{\rm A}=\bigl[sy(Z_{\rm A}^{1}),sy(Z_{\rm A}^{2})\bigr]$, where $sy(\cdot)$ is a linear function prescribed by a low density parity check (LDPC) code~\cite{Luby1998a_,Pearson2004} for a prefixed QBER, and all of them jointly reconstruct $sy_{\rm A}$ via the Reconstruct protocol of a VSS scheme (see the toolbox section in this Supplemental Material). Coming next, all four units perform a RBS generation protocol to select the two-universal hash functions $h_{\rm EV}$ (to be used for error verification) and $h_{\rm PA}$ (to be used for privacy amplification). The output length of $h_{\rm EV}$ is set to $t_{\rm EV}=64$ bits and the output length of $h_{\rm PA}$ is set to $l=\min\{l_{1},l_{2}\}$ (see the next section in this Supplemental Material for a justification of this). Then, all four units individually compute their shares of the EV tag, $h_{\rm EV}(Z_{\rm A})$, and later on reconstruct it via the Reconstruct protocol of a VSS scheme. In a single communication to $\textrm{CP}_{\rm B}$, every unit in $\mathcal{A}$ send:
	\begin{enumerate}
		\item{The syndrome information $sy_{\rm A}$, a description of $h_{\rm EV}$ and the EV tag $h_{\rm EV}({Z}_{\rm A})$.}
		\item{The description of $h_{\rm PA}$.}
	\end{enumerate}
	$\textrm{CP}_{\textrm{B}}$ decides on each item via MV. Using $sy_{\rm A}=\bigl[sy(Z_{\rm A}^{1}),sy(Z_{\rm A}^{2})\bigr]$ and $\textrm{Z}_{\rm B}=\textrm{Z}_{\rm B}^{1}\textrm{Z}_{\rm B}^{2}$, $\textrm{CP}_{\textrm{B}}$ implements the decoding scheme of the LDPC code ---based on a log-likelihood ratio belief propagation algorithm--- to construct its reconciled string $\hat{Z}_{\rm B}=\hat{Z}_{\rm B}^{1}\hat{Z}_{\rm B}^{2}$. Importantly, the decoding scheme is applied separately for $j=1$ and $j=2$ from the respective syndromes. Then, $\textrm{CP}_{\textrm{B}}$ computes the EV tag $h_{\rm EV}(\hat{Z}_{\rm B})$. If $h_{\rm EV}(\hat{Z}_{\rm B})\neq{}h_{\rm EV}({Z}_{\rm A})$ it aborts the protocol.
	\item\textbf{Privacy amplification.}
	In case of no abortion, $\textrm{CP}_{\textrm{B}}$ computes Bob's final key $S_{\rm B}=h_{\rm PA}(\hat{Z}_{\rm B})$. Similarly, if no abortion is notified, all of Alice's units compute their shares of Alice's final key, $S_{\rm A}=h_{\rm PA}({Z}_{\rm A})$, by applying $h_{\rm PA}$ to their shares of ${Z}_{\rm A}$.
\end{enumerate}
\section{Secret key length and parameter estimation}\label{sec:PE}
A full derivation of the extractable secret key length in a slightly more general protocol than the one considered above is given in~\cite{Victor}, and the bulk of the mathematical derivation presented there is standard in QKD. To be precise, it builds up on the leftover-hash lemma against quantum side information~\cite{Tomamichel} and the uncertainty relation for smooth entropies~\cite{Tomamichel_2}, together with two chain inequalities respectively presented in~\cite{Renner} and~\cite{Vitanov}. For the protocol above, it follows that one can extract $l$ $\epsilon_{\rm sec}$-secret, $\epsilon_{\rm cor}$-correct key bits via privacy amplification with 2-universal hashing, for~\cite{Victor}

\begin{equation}\label{secrecy}
l=32\times{}\left\lfloor{\frac{\min\{l_{1},l_{2}\}}{32}}\right\rfloor\hspace{.5cm}\textrm{and for all}\hspace{.5cm}\epsilon_{\rm sec}\geq{}\hat{\epsilon}_{\rm sec}+\epsilon_{\rm AU},\hspace{.5cm}\epsilon_{\rm cor}\geq{}\hat{\epsilon}_{\rm cor}+\epsilon_{\rm AU}
\end{equation}
where $\hat{\epsilon}_{\rm sec}=2\varepsilon+\delta+\epsilon_{\rm PA}$, $\epsilon_{\rm AU}$ is the total error probability of the authentication (computed in the final section), $\hat{\epsilon}_{\rm cor}=2^{-t_{\rm EV}+1}\times{}|Z_{\rm A}|$ and we shortened the key length to be a multiple of 32 because it is a restriction of our PA scheme. Note that $l_{j}$, $\epsilon_{\rm PA}$ and $\delta$ were already presented in Eq.~(\ref{l_j_}). As for $\varepsilon$, it denotes the error probability of the parameter estimation step. Precisely, if we denote the pair index corresponding to the honest QKD pair by $j=\mathcal{J}$, $\varepsilon$ is upper bounded by the sum of the error probabilities of the one-sided statistical bounds $S_{11,\rm Z}^{\mathcal{J},\rm L}$ and $\phi_{11,\rm Z}^{\mathcal{J}, \rm U}$. The parties pre-agree on a common parameter estimation error that applies to both $j=1$ and $j=2$. Therefore, $\varepsilon$ matches this pre-agreed value.

Although, as stated above, the reader is referred to~\cite{Victor} for a step-by-step derivation of Eqs.~(\ref{l_j_}) and (\ref{secrecy}), the idea behind the structure $l=\min\{l_{1},l_{2}\}$ in Eq.~(\ref{secrecy}) is that, in the presence of an honest QKD pair which certainly delivers faithful protocol information for the parameter estimation, $\min\{l_{1},l_{2}\}$ is guaranteed to be a valid lower bound on $l_{\mathcal{J}}$, which indeed defines the key length extractable via PA.\\

Regarding the parameter estimation procedure, we use the decoy-state bounds derived in~\cite{Curty2_}. Importantly though, while the protocol in~\cite{Curty2_} uses three intensity settings for each basis (Z and X), we recall that our experiment follows the efficient MDI-QKD scheme proposed in~\cite{X.B.Wang_}. Therefore, we combine the bounds in~\cite{Curty2_} with various standard results in large deviation theory ---the Chernoff bound (see for instance~\cite{Mitzenmacher}), the Serfling inequality~\cite{Serfling} and the inverse Chernoff bound~\cite{Zhang}--- to complete the parameter estimation. In fact, very minor changes allow to adapt the explicit formulas for $S_{11,\rm Z}^{j,\rm L}$, $\phi_{11,\rm Z}^{j,\rm U}$ and $\varepsilon$ in~\cite{Victor} to the reported experiment.

\section{Experimental data}
The quantum communication run for 16000 seconds for each QKD pair, with an emulated attenuation of 12 \si{\decibel} on each side. We choose the decoy intensity settings as $\{\lambda, \mu, \nu, \omega\}=\{0.23,0.23,0.047,0.005\}$, $q_{\rm X}=0.41$ and $\{p_{\mu},p_{\nu},p_{\omega}\}=\{0.2,0.6,0.2\} $, which are near to the optimal settings according to a typical channel model. For example, we select $\lambda=\mu$ for simplicity, as the resulting performance is very similar to that corresponding to the optimal intensities. Tabs.~\ref{tab:rectilinear},~\ref{tab:CX} and~\ref{tab:EX} show the measured counts and the error rate of the Z (X) basis for both QKD pairs.
\begin{table}[htb]
	\caption{Measured counts and errors in the rectilinear basis\label{tab:rectilinear}}
	\begin{tabular}{|@{\hspace{0.2cm}} c @{\hspace{0.2cm}} |@{\hspace{0.5cm}} c @{\hspace{0.5cm}}|@{\hspace{0.5cm}} c @{\hspace{0.5cm}}|}
		\hline \hline
		$ j $ & 1 & 2\\
		\hline
		$ \left|\mathcal{Z}_{j}\right|$    & 109759094   & 111149334\\			\hline
		$e_{j}/\left|\mathcal{Z}_{j}\right| $    &2.30\%  &	2.13\%\\	
		\hline \hline		
	\end{tabular}
\end{table}
\begin{table}[htb]
	\caption{Measured counts in the diagonal basis\label{tab:CX}}
	\begin{ruledtabular}
		\begin{tabular}{|c|c|c|c|c|c|c|}
			&\multicolumn{3}{c|}{$ j=1 $}& \multicolumn{3}{c|}{$ j=2 $}	\\
			\hline
			$ \left|\mathcal{X}_{j}^{a, b}\right|$ &	$ \mu $     &$ \nu $     &$ \omega $&	$ \mu $     &$ \nu $     &$ \omega $\\
			\hline
			$ \mu $ &	4124600     &4497145     &1093752	&	4124576    &4492161     &1089912\\
			\hline
			$ \nu $ &	4465220     &1550736      &163836	&4470030     &1544726      &160096\\
			\hline
			$ \omega $ &	1075116      &157763        &1997	&1079895      &159422        &2181
		\end{tabular}
	\end{ruledtabular}
\end{table}

\begin{table}[htb]
	\caption{Measured QBER in the diagonal basis\label{tab:EX}}
	\begin{ruledtabular}
		\begin{tabular}{|c|c|c|c|c|c|c|}
			&\multicolumn{3}{c|}{$ j=1 $}& \multicolumn{3}{c|}{$ j=2 $}	\\
			\hline
			$ e^{a,b}_{j}/\left|\mathcal{X}_{j}^{a, b}\right| $ &	$ \mu $     &$ \nu $     &$ \omega $&	$ \mu $     &$ \nu $     &$ \omega $\\
			\hline
			$ \mu $ &	27.33\%     &37.81\%     &48.79\%	&	26.85\%    &37.13\%     &48.13\%\\
			\hline
			$ \nu $ &	37.42\%     &28.29\%      &44.82\%	&37.59\%     &27.91\%      &43.15\%\\
			\hline
			$ \omega $ &	48.51\%      &43.56\%        &37.36\%	&48.65\%      &44.53\%        &36.50\%
		\end{tabular}
	\end{ruledtabular}
\end{table}

Using $ \left|\mathcal{Z}_j\right|,\left|\mathcal{X}_{j}^{a, b}\right| $ and $ e_j^{a,b} $, the parameter estimation procedure described in the previous section outputs $ S_{11,Z}^{j,\rm L} $ and $ \phi_{11,Z}^{j,\rm U} $. Here we use a code rate $ R $ of 81\% for the EC (suitable for a $\rm QBER\approx{}2.5\%$). In principle, this leads to a syndrome of $ \lambda_{\mathrm{EC}}^{j}=|sy(Z_{\rm A}^{j})|=\lceil{(1-R)\cdot \left|\mathcal{Z}_j\right|}\rceil$ bits in Eq.~(\ref{l_j_}). However, EC requires to divide the sifted key in blocks of length $M=2^{16}$, and it further requires the syndrome of each block to be a multiple of $8$. Therefore, we pad both sifted keys and the syndromes of their blocks with zeros in order to match these conditions. As a consequence, one must replace $\left|\mathcal{Z}_j\right|$ by $M\left\lceil{\left|\mathcal{Z}_j\right|/M}\right\rceil$ and $ \lambda_{\mathrm{EC}}^{j}$ by
$\left\lceil{\left|\mathcal{Z}_j\right|/M}\right\rceil\times{}8\left\lceil{(1-R)M/8}\right\rceil$. Finally, $ l_j $ is computed using Eq.~(\ref{l_j_}) with $\delta=\epsilon_{\rm PA}=\hat{\epsilon}_{\rm sec}/46$. The denominator 46 follows because we set a common value for every contribution to $\hat{\epsilon}_{\rm sec}=2\varepsilon+\delta+\epsilon_{\rm PA}$, and $\varepsilon$ contains 22 parameter estimation error terms~\cite{Victor}. The results are shown in Tab.~\ref{tab:result}.\\

\begin{table}[!htbp]
	\centering
	\caption{Parameter estimation results and extractable secret key length for each QKD pair. $S_{11,\rm Z}^{j,\rm L}$ ($\phi_{11,\rm Z}^{j,\rm U}$) denotes a lower (upper) bound on the number of single-photon successes (single-photon phase-error rate) in the sifted key, and $\hat{\epsilon}_{\rm sec}$ ($\hat{\epsilon}_{\rm cor}$) is the secrecy (correctness) parameter of the final keys disregarding the overall authentication error probability, $\epsilon_{\rm AU}$.}
	\label{tab:result}
	\begin{tabular}{|@{\hspace{0.5cm}} c @{\hspace{0.5cm}}|@{\hspace{0.5cm}} c @{\hspace{0.5cm}}|@{\hspace{0.5cm}} c @{\hspace{0.5cm}}|@{\hspace{0.5cm}} c @{\hspace{0.5cm}}|@{\hspace{0.5cm}} c @{\hspace{0.5cm}}|@{\hspace{0.5cm}} c @{\hspace{0.5cm}}|}
		\hline \hline
		$j$   & $ S_{11,Z}^{j,\rm L} $ & $ \phi_{11,Z}^{j,\rm U} $  & $ \hat{\epsilon}_{\rm sec} $ &$ \hat{\epsilon}_{\rm cor} $ & $ l_j $     \\
		\hline
		1 & 50654051 & 0.1105 & \multirow{2}{*}{$ 10^{-8} $}& \multirow{2}{*}{$ 2.4\times{}10^{-11} $}& 4386592 \\ 
		2 & 50887187 & 0.1075 & & &4694048 \\
		\hline \hline
	\end{tabular}
\end{table}
We remark that the total error probability of the authentication of the classical communications, $\epsilon_{\rm AU}$ (quantified in the last section of this Supplemental Material), contributes to both $\epsilon_{\rm sec}$ and $\epsilon_{\rm cor}$, as stated in Eq.~(\ref{secrecy}). As we shall see below, while $\epsilon_{\rm sec}=\hat{\epsilon}_{\rm sec}+\epsilon_{\rm AU}\approx{}\hat{\epsilon}_{\rm sec}$, $\epsilon_{\rm cor}=\hat{\epsilon}_{\rm cor}+\epsilon_{\rm AU}\approx{}\epsilon_{\rm AU}$. That is to say, the correctness parameter is dominated by the authentication term.

Finally, the experimental secret key rate is computed as
\begin{equation}
K=\frac{l-l_{\rm AU}}{2N},
\end{equation}
where $l_{\rm AU}$ is the secret key length consumed for the authentication of the classical communications (computed in the last Section of this Supplemental Material). Also, we recall that $l$ is the extractable key length (given in Eq.~(\ref{secrecy})) and $N=2\times{}10^{13}$ is the number of signals transmitted per QKD session.
\section{Data post-processing}
In order to verify the BSM events correctly, the timing of the detection events is registered with a synchronization signal from the clock board. For this purpose, we use a Swabian Instrument Time Tagger Ultra. This allows to extract the coincidence events with a coincidence window of 400 \si{\pico\second}, which covers more than a 99.7\% of the Gaussian pulse while maintaining the QBER relatively low.

After the QKD session and the distribution of the necessary data, the multiparty data post-processing in our protocol has the following steps in order: sifting, parameter estimation, random permutation, information reconciliation (IR), error verification (EV) and privacy amplification (PA). The overall procedure is similar to a typical one in~\cite{Fung2010}, although some techniques are drawn from classical secure multiparty computation (MPC).

In our implementation, we use a particular class of two-universal hash functions ---Toeplitz matrices--- to perform EV, PA and authentication, although there exist differences in the construction of the matrices and the source of the random bits in each case.

On the one hand, the $\mathrm{CP}_{\mathrm{A}_{l}}$ use an RBS generation protocol (see the toolbox section in this Supplemental Material) to randomly select the EV Toeplitz matrix, $h_{\rm EV}$, and the PA Toeplitz matrix, $h_{\rm PA}$. In the first place, $h_{\rm EV}$ is constructed following a linear feedback shift register (LFSR) technique~\cite{Krawczyk1994}. With this method, an EV tag of pre-agreed length $t_{\rm EV}=64$ bits is computed with a matrix $h_{\rm EV}$ specified with 128 random bits. In the second place, $h_{\rm PA}$ is a fully random Toeplitz matrix. Its specification requires $(n+l-1)$ random bits, where $ n $ is the input length of the PA, which is 220987392 bits (length of the concatenation of the padded sifted keys) and $ l $ is the output length of the PA determined by Eq.~(\ref{secrecy}), 4386592 bits.

On the other hand, the authentication scheme is explained in detail in Sec.~\ref{Authentication} of this Supplemental Material. The scheme requires every unit in $\mathcal{A}$ to pre-share a pool of secret key bits with $\mathrm{CP}_{\mathrm{B}}$, and these dedicated pools are the source of randomness required for the authentication of the classical messages. In fact, the authentication of any given classical message is formally identical to EV, in the sense that the authentication tags (whose length is also fixed to 64 bits) are also computed using LFSR-based Toeplitz matrices. We remark though that such matrices are built drawing secret bits from the dedicated key pools.

As stated in Section III of this Supplemental Material, prior to IR, each sifted key is divided in blocks of $2^{16}$ bits (padding the last block with zeros) and randomly permuted to prevent burst errors that may cause the failure of the IR. We remark that the permutation matrix is publicly selected a priori. Coming next, IR is also performed blockwise. In the first place, EC is conducted using a LDPC code, whose syndrome information is prescribed based on irregular graphs~\cite{Luby1998a_}. The length of the syndrome for each block is determined by the code rate $R$ of the LDPC code, and the syndromes are padded with zeros in order to contain an integer number of bytes. In particular, we select a code rate $R=81\%$, which suffices to correct a prefixed threshold QBER of 2.5\%. After obtaining the syndromes, CP$ _\mathrm{B} $ implements the decoding scheme of the LDPC code ---based on a log-likelihood ratio belief propagation algorithm~\cite{Chen2005_}--- to construct its reconciled string $\hat{Z}_{\rm B}=[\hat{Z}_{\rm B}^{1},\hat{Z}_{\rm B}^{2}]$. We note, however, that the decoding algorithm could also be implemented in a distributed way. That is, $\textrm{CP}_{\rm B}$ could perform the syndrome calculation and Alice's units could execute the decoding algorithm. In fact, we also demonstrated this alternative experimentally, as explained in the next section.
\section{Distributed error correction decoding}
Despite EC is performed by Bob in the actual experiment, for completeness we have separately implemented the decoding algorithm of EC also in a distributed way using Alice's CP units. Here, we briefly explain how this alternative decoding works. First of all, since EC is implemented independently for each pair of sifted keys, for the explanation it suffices to focus on, say, the first pair, $(Z_{\rm A}^{1},Z_{\rm B}^{1})$. Let $\hat{e}^{1}$ be the error pattern between $Z_{\rm A}^{1}$ and $Z_{\rm B}^{1}$, such that $Z_{\rm A}^{1}\oplus{}Z_{\rm B}^{1}=\hat{e}^{1}$. In the original approach described in the main text, the $\mathrm{CP}_{\mathrm{A}_{l}}$ compute $sy(Z_{\rm A}^{1})$ distributedly and send it to CP$_{\rm B}$, which applies the decoding algorithm of the EC protocol to identify the most likely string $\hat{e}^{1}$ from the pair of inputs $Z_{\rm B}^{1}$ and $sy(Z_{\rm A}^{1})$. We remark that, since CP$_{\rm B}$ is assumed to be honest, the fact that it knows $Z_{\rm B}$ does not compromise security. Nevertheless, if the EC decoding is implemented at Alice's side instead, the decoding algorithm should be implemented in such a way that none of her units can learn $Z_{\rm A}^{1}$. For this, the procedure must be slightly modified in order not to compromise security. Precisely, using the fact that
\begin{equation}
Z_{\rm A}^{1}\oplus{}Z_{\rm B}^{1}=\hat{e}^{1}\iff{}Z_{\mathrm{A},4}^{1}\oplus\left(Z_{\mathrm{A},1}^{1}\oplus{}Z_{\mathrm{A},2}^{1}\oplus{}Z_{\mathrm{A},3}^{1}\oplus{}Z_{\rm B}^{1}\right)=\hat{e}^{1},
\end{equation}
where we recall that $Z_{\textrm{A},k}^{1}$ is the $k$-th share of $Z_{\textrm{A}}^{1}$, it follows that the output of the decoding algorithm for the inputs $Z_{\rm A}^{1}$ and $sy(Z_{\rm B}^{1})$ is the same as for the inputs $Z_{\mathrm{A},4}^{1}$ and $sy(Z_{\mathrm{A},1}^{1}\oplus{}Z_{\mathrm{A},2}^{1}\oplus{}Z_{\mathrm{A},3}^{1}\oplus{}Z_{\rm B}^{1})$, with the crucial difference that the second approach does not compromise the secrecy of $Z_{\rm A}^{1}$. This is so because the $\mathrm{CP}_{\mathrm{A}_{l}}\in\mathcal{A}$ already know the fourth share $Z_{\mathrm{A},4}^{1}$ in advance (first input), and the information revealed by $sy(Z_{\mathrm{A},1}^{1}\oplus{}Z_{\mathrm{A},2}^{1}\oplus{}Z_{\mathrm{A},3}^{1}\oplus{}Z_{\rm B}^{1})$ (second input) is already accounted for in the secret key length formula. In order to implement this second approach, CP$_{\rm B}$ computes $sy(Z_{\rm B}^{1})$ and sends it to the $\mathrm{CP}_{\mathrm{A}_{l}}\in\mathcal{A}$. Then, the latter individually compute $sy(Z_{\mathrm{A},1}^{1}\oplus{}Z_{\mathrm{A},2}^{1}\oplus{}Z_{\mathrm{A},3}^{1}\oplus{}Z_{\rm B}^{1})$ by bitwise XOR-ing $sy(Z_{\mathrm{A},1}^{1})$, $sy(Z_{\mathrm{A},2}^{1})$, $sy(Z_{\mathrm{A},3}^{1})$ and $sy(Z_{\rm B}^{1})$. This task requires that, for $k=1,2,3$, $\mathrm{CP}_{\mathrm{A}_{k}}$ learns its missing syndrome share, $sy(Z_{\rm A}^{k})$. For this, $\mathrm{CP}_{\mathrm{A}_{k}}$ requests the share to the remaining three units (who certainly hold it) and applies MV. Finally, the $\mathrm{CP}_{\mathrm{A}_{l}}\in\mathcal{A}$ individually execute the decoding algorithm.
\section{Secure multiparty computation toolbox}
Below we present the secure MPC tools that we use in the post-processing scheme of the experimental MDI-QKD implementation. For convenience, these tools are not described in full generality, but they are particularised for the scenario in hand, where Alice holds four CP units and at most one of them might be corrupted by an active adversary. For more general descriptions, see for instance~\cite{Maurer_,Cramer_}.
\subsection{Share protocol of a VSS scheme}
Here, we explain the Share protocol that $\textrm{QKD}_{\textrm{A}_{j}}$ uses to distribute shares of ${r}^{j}|_{\mathcal{I}_j,\rm Z}$.
\begin{enumerate}
	\item{$\textrm{QKD}_{\textrm{A}_{j}}$ generates three random bit strings of length $\bigl|{r}^{j}|_{\mathcal{I}_j,\rm Z}\bigr|$, denoted by $S_{1}^{j}$, $S_{2}^{j}$ and $S_{3}^{j}$. Then, it generates a fourth string defined as $S_{4}^{j}=S_{1}^{j}\oplus{}S_{2}^{j}\oplus{}S_{3}^{j}\oplus{}{r}^{j}|_{\mathcal{I}_j,\rm Z}$}.
	\item{For $k=1,...,4$, $\textrm{QKD}_{\textrm{A}_{j}}$ sends $S_{k}^{j}$ to every $\textrm{CP}_{\textrm{A}_{l}}$ distinct from $\textrm{CP}_{\textrm{A}_{k}}$. Thus, by construction, each unit will only be missing one share of ${r}^{j}|_{\mathcal{I}_j,\rm Z}$. If a unit does not receive a prescribed share, it takes this share to be a null bit string by default.}
	\item{Let $\sigma_{k}$ denote the set of units that hold the $k$-th share of ${r}^{j}|_{\mathcal{I}_j,\rm Z}$, $S_{k}^{j}$. As an example, $\sigma_{1}=\{\textrm{CP}_{\textrm{A}_{2}},\textrm{CP}_{\textrm{A}_{3}},\textrm{CP}_{\textrm{A}_{4}}\}$. For $k=1,...,4$, all the units in $\sigma_{k}$ send their copies of $S_{k}^{j}$ to each other, in order to check that they are indeed equal. If a unit finds an inconsistency, it raises a complaint through a (possibly simulated) broadcast channel and $\textrm{QKD}_{\textrm{A}_{j}}$ broadcasts $S_{k}^{j}$ to all four units.}
\end{enumerate}
\subsection{Reconstruct protocol of a VSS scheme}
Here, we explain how the Reconstruct protocol of a VSS scheme runs for the reconstruction of the syndrome information $sy_{\rm A}=\bigl[sy(Z_{\rm A}^{1}),sy(Z_{\rm A}^{2})\bigr]$ by Alice's units. The reconstruction of the EV tag $h_{\rm EV}(Z_{\rm A})$ is performed identically.

Let us denote the $k$-th share of
the sifted key $Z_{\rm A}^{j}$ by $Z_{\textrm{A},k}^{j}$, which is obtained by sifting the $k$-th share of ${r}^{j}|_{\mathcal{I}_j,\rm Z}$. It follows that $Z_{\rm A}^{j}=Z_{\textrm{A},1}^{j}\oplus{}Z_{\textrm{A},2}^{j}\oplus{}Z_{\textrm{A},3}^{j}\oplus{}Z_{\textrm{A},4}^{j}$. Then, the $k$-th share of $sy(Z_{\rm A}^{j})$ is simply given by $sy(Z_{\textrm{A},k}^{j})$, and the linearity of the syndrome function $sy(\cdot)$ implies that
\begin{equation}
sy(Z_{\rm A}^{j})=\bigoplus_{k=1}^{4}sy\left(Z_{\textrm{A},k}^{j}\right).
\end{equation}
Moreover, by construction $sy(Z_{\textrm{A},k}^{j})$ is the only share missing for $\textrm{CP}_{\textrm{A}_{k}}$, for all $k=1,\ldots,4$. For each $j$, the Reconstruct protocol of $sy(Z_{\rm A}^{j})$ goes as follows.
\begin{enumerate}
	\item{For $k=1,...,4$, each of the three units in $\sigma_{k}$ sends its copy of $sy(Z_{\textrm{A},k}^{j})$ to $\textrm{CP}_{\textrm{A}_{k}}$.}
	\item{For each $k$, $\textrm{CP}_{\textrm{A}_{k}}$ uses MV to decide on $sy(Z_{\textrm{A},k}^{j})$ and later on bitwise XORs all the shares to reconstruct $sy(Z_{\rm A}^{j})$.}
\end{enumerate}

Crucially, the VSS scheme presented above (including both protocols, Share and Reconstruct) enables information-theoretically secure MPC and is due to~\cite{Maurer_}.
\subsection{Random bit string (RBS) generation protocol}
Below we present a simple procedure used by Alice's CP units to create the random bit-strings that specify the hash functions $h_{\rm EV}$ and $h_{\rm PA}$, in the presence of a corrupted unit. Remarkably, the protocol does not require any synchrony assumption among the participating units.
\begin{enumerate}
	\item{$\textrm{CP}_{\textrm{A}_{1}}$ ($\textrm{CP}_{\textrm{A}_{2}}$) generates a random bit string $R_{1}$ ($R_{2}$) of a prefixed length and sends it to $\textrm{CP}_{\textrm{A}_{3}}$. If $\textrm{CP}_{\textrm{A}_{3}}$ does not receive, say $R_{j}$, it raises a complaint through a broadcast channel and $\textrm{CP}_{\textrm{A}_{j}}$ broadcasts this string.}
	\item{$\textrm{CP}_{\textrm{A}_{3}}$ computes $R=R_{1}\oplus{R_{2}}$ and sends it back to  $\textrm{CP}_{\textrm{A}_{1}}$ and $\textrm{CP}_{\textrm{A}_{2}}$. If a unit does not receive $R$, it raises a complaint through a broadcast channel and $\textrm{CP}_{\textrm{A}_{3}}$ broadcasts this string.}
	\item{$\textrm{CP}_{\textrm{A}_{1}}$ ($\textrm{CP}_{\textrm{A}_{2}}$) computes ${R}_{2}=R\oplus{R}_{1}$ (${R}_{1}=R\oplus{R}_{2}$) and sends it to $\textrm{CP}_{\textrm{A}_{2}}$ ($\textrm{CP}_{\textrm{A}_{1}}$). $\textrm{CP}_{\textrm{A}_{2}}$ ($\textrm{CP}_{\textrm{A}_{1}}$) checks that the string received from $\textrm{CP}_{\textrm{A}_{1}}$ ($\textrm{CP}_{\textrm{A}_{2}}$) matches the one it sent to $\textrm{CP}_{\textrm{A}_{3}}$ in step 1. Otherwise, it raises a complaint and $\textrm{CP}_{\textrm{A}_{3}}$ broadcasts $R$. In this case, $\textrm{CP}_{\textrm{A}_{1}}$ and $\textrm{CP}_{\textrm{A}_{2}}$ redo step 3.}
\end{enumerate}
Upon completion of steps 1 to 3, two honest units in $\mathcal{A}$ hold a common random string $R$. Then, the three of them send $R$ to $\textrm{CP}_{\textrm{A}_{4}}$, which applies MV.

Let us briefly discuss the security of the protocol above. Due to the symmetric roles of $\textrm{CP}_{\textrm{A}_{1}}$ and $\textrm{CP}_{\textrm{A}_{2}}$, it suffices to address the case where, say $\textrm{CP}_{\textrm{A}_{1}}$ is dishonest, and the case where $\textrm{CP}_{\textrm{A}_{3}}$ is dishonest. Let us first consider that $\textrm{CP}_{\textrm{A}_{1}}$ is dishonest. Then, $\textrm{CP}_{\textrm{A}_{3}}$ is honest, which forces $\textrm{CP}_{\textrm{A}_{1}}$ to send him $R_{1}$ before knowing $R_{2}$. Since $R_{2}$ is random due to the honesty of $\textrm{CP}_{\textrm{A}_{2}}$ and $R_{1}$ is uncorrelated to $R_{2}$, the string $R=R_{1}\oplus{}R_{2}$ delivered by $\textrm{CP}_{\textrm{A}_{3}}$ in step 2 is random. On the other hand, let us assume instead that $\textrm{CP}_{\textrm{A}_{3}}$ is dishonest. Then, $\textrm{CP}_{\textrm{A}_{1}}$ and $\textrm{CP}_{\textrm{A}_{2}}$ are honest, in such a way that they generate random strings $R_{1}$ and $R_{2}$, and either $\textrm{CP}_{\textrm{A}_{3}}$ delivers $R=R_{1}\oplus{}R_{2}$ to both $\textrm{CP}_{\textrm{A}_{1}}$ and $\textrm{CP}_{\textrm{A}_{2}}$ in step 3, or at least one of them will complain, in which case $\textrm{CP}_{\textrm{A}_{3}}$ is forced to broadcast $R$. Step 3 is repeated until the broadcasted value actually matches $R_{1}\oplus{}R_{2}$. Therefore, upon completion of the protocol, both $\textrm{CP}_{\textrm{A}_{1}}$ and $\textrm{CP}_{\textrm{A}_{2}}$ end up with a common random string $R_{1}\oplus{}R_{2}$.

We remark that, although the protocol above is tailored for the scenario where only one CP unit might be corrupted and at least three CP units are available, a more general protocol based on VSS can be found in~\cite{Curty_}. This alternative protocol is guaranteed to succeed as long as the number of honest CP units is larger than two thirds of the total number of units.

\section{Authentication cost}\label{Authentication}
To start up with, we list the classical messages of the protocol and specify their redundancy (in brackets):
\begin{enumerate}
	\item{$m_{\rm B}^{1}=\left[b^{1}|_{\mathcal{I}_{1}},{r'}^{1}|_{\mathcal{I}_1,\rm X}\right]$ and $m_{\rm B}^{2}=\left[b^{2}|_{\mathcal{I}_{2}},{r'}^{2}|_{\mathcal{I}_2,\rm X}\right]$ (each of them sent by $\textrm{CP}_{\rm B}$ to all three $\mathrm{CP}_{\mathrm{A}_{l}}\in\mathcal{A}$).}
	\item{$m_{\rm A}^{1,\rm SIFT}=\mathcal{Z}_{1}$ and $m_{\rm A}^{2,\rm SIFT}=\mathcal{Z}_{2}$ (each of them sent by all three $\mathrm{CP}_{\mathrm{A}_{l}}\in\mathcal{A}$ to $\textrm{CP}_{\rm B}$).}
	\item{$m_{\rm A}^{\textrm{IR,PA}}=\left[sy(Z_{\rm A}^{1}),sy(Z_{\rm A}^{2}),h_{\rm EV}(Z_{\rm A}),h_{\rm EV},h_{\rm PA}\right]$ (sent by all three $\mathrm{CP}_{\mathrm{A}_{l}}\in\mathcal{A}$ to $\textrm{CP}_{\rm B}$).}
\end{enumerate}
The authentication scheme we follow is described in~\cite{Fung2010}. Let us focus on any given $\mathrm{CP}_{\mathrm{A}_{l}}\in\mathcal{A}$. For the authentication of a message $m$ (of arbitrary size) both $\mathrm{CP}_{\mathrm{A}_{l}}$ and $\mathrm{CP}_{\mathrm{B}}$ locally generate a common LFSR-based Toeplitz matrix, $T$, by drawing secret bits from their pre-shared key pool (see Section IV of this Supplemental Information). The sending unit computes the tag $t=T\times{m}$ and encrypts it later on using the one-time pad. The encrypted tag is then attached to the message, and the overall key cost reduces to that of encrypting the tag, \textit{i.e.}, $|t|$ bits. That is to say, the secret bits consumed for the construction of Toeplitz matrices remain secure and can be reallocated in the key pool.

For practical purposes, the length of every authentication tag is fixed to 64 bits in the experiment. Moreover, from the list above, there exist five different messages: $m_{\rm B}^{1}$, $m_{\rm B}^{2}$, $m_{\rm A}^{1,\rm SIFT}$, $m_{\rm A}^{2,\rm SIFT}$ and $m_{\rm A}^{\textrm{IR,PA}}$. Since each of them is redundantly sent three times, it follows that, in total, authentication consumes $l_{\rm AU}=64\times{5}\times{3}=960$ bits.

In addition, the error probability of the authentication scheme~\cite{Fung2010} is given by
\begin{equation}
\gamma_{\rm AU}=|m|2^{-|t|+1}.
\end{equation}
Therefore, the overall error probability $\epsilon_{\rm AU}$ of the authentication satisfies
\begin{equation}\label{cost}
\epsilon_{\rm AU}\leq{}\frac{3}{2^{63}}\left[|m_{\rm B}^{1}|+|m_{\rm B}^{2}|+\left|m_{\rm A}^{1,\rm SIFT}\right|+\left|m_{\rm A}^{2,\rm SIFT}\right|+\left|m_{\rm A}^{\textrm{IR,PA}}\right|\right].
\end{equation}
In fact, only the wrong authentication of those messages both sent and received by honest CP units compromises the secrecy, meaning that the above formula can be made tighter. Precisely, the prefactor $3\times{2^{-63}}$ in Eq.~(\ref{cost}) can be simplified to ${2^{-62}}$. Nevertheless, the difference is not relevant due to the smallness of $\epsilon_{\rm AU}$ in either case. To finish with, we list the precise message lengths below:
$|m_{\rm B}^{1}|=578333424$ bits, 
$|m_{\rm B}^{2}|=596429040$ bits,
$\left|m_{\rm A}^{1,\rm SIFT}\right|=578333424$ bits,
$\left|m_{\rm A}^{2,\rm SIFT}\right|=596429040$ bits, and
$\left|m_{\rm A}^{\textrm{IR,PA}}\right|=267375807$ bits. For the last item, we used the fact that $ |sy(Z_{\rm A}^{1})|=\lambda_{\mathrm{EC}}^{1}=20863800$ bits, $|sy(Z_{\rm A}^{2})|=\lambda_{\mathrm{EC}}^{2}=21137832$ bits, $|h_{\rm EV}({Z}_{\rm A})|=t_{\rm EV}=64$ bits, $|h_{\rm EV}|=128$ bits and $|h_{\rm PA}|=225373983$ bits. Putting it all together, it follows that $\epsilon_{\rm AU}\approx{5.7\times{}10^{-10}}$.

\end{document}